# Algorithms for Computing in Fog Systems: principles, algorithms, and Challenges

Nikheel Soni[1,2], Reza Malekian[2,3*], and Dijana Capeska Bogatinoska[4]

[1]Amazon Web Services, Cape Town, 8001, South Africa
[2]Department of Electrical, Electronic and Computer Engineering, University of Pretoria, Pretoria, 0002, South Africa
[3]Department of Computer Science and Media Technology, Malmö University, Malmö, 20506, Sweden
[4]University of Information Science and Technology "St. Paul the Apostle", Ohrid, Macedonia
nikheel@amazon.com, reza.malekian@ieee.org (corresponding author), dijana.c.bogatinoska@uist.edu.mk

*Abstract* - **Fog computing is an architecture that is used to distribute resources such as computing, storage, and memory closer to end-user to improve applications and service deployment. The idea behind fog computing is to improve cloud computing and IoT infrastructures by reducing compute power, network bandwidth, and latency as well as storage requirements. This paper presents an overview of what fog computing is, related concepts, algorithms that are present to improve fog computing infrastructure as well as challenges that exist. This paper shows that there is a great advantage of using fog computing to support cloud and IoT systems.**

## I. Introduction

Fog computing is an architecture that is used to distribute resources closer to users, at the edge of the network, to improve the overall performance of applications and services [1], [2]. Fog computing is not designed to replace cloud computing, but to complement it by distributing computing, communication, control, and storage elements closer to the end-users [1]. The use of fog computing can be seen in various applications and services that do not work well in the cloud paradigm which include applications that are latency dependent such as gaming, geo-distributed such as pipeline monitoring, fast mobile applications such as connected vehicles and large-scale distribution such as smart grid [2]. Other applications where fog computing can prove to be highly effective include augmented reality, content delivery and caching and analytics on big data [1], [2]. As latency is a topic that is widely discussed, latency reduction does not occur instantly and depends on where application components are placed and is not critical for every application [3]. Environments in which fog computing architecture is implemented need to be thoroughly evaluated as an incorrect assessment can result in an increase in factors such as costs, security risks and processing time.

For instance, if an Internet of Things (IoT) environment is constantly changing and developing, implementation of fog can prove to be beneficial, as it will enable easy integration for any changes made. If an IoT environment does not change at all, the implementation of fog could be redundant. Implementation of fog should be considered when many devices are generating data, there is a high priority in obtaining a result from data that is generated and in areas where there is intermittent access to the cloud [9], Fig.1. Making use of fog computing architecture has numerous advantages. Advantages include: security, cognition, agility, latency, and efficiency [1]. Security is improved as data needs to travel a smaller distance which makes it less susceptible to attacks, fog nodes acting as proxies used to deploy certain security functions [1], keeping sensitive data is contained inside the network [4], act as the first line of access control and traffic encryption and provide data integrity to mention a few [1]. Cognition is created as fog devices are closer to the user, therefore, resulting in a better understanding of the user's requirements for an application [1]. The architecture is agile as it allows for instantaneous scaling concerning both hardware and software instances [1]. By reducing latency, time-constrained applications can be performed much faster to obtain results. Lastly, efficiency is improved through all the mentioned advantages by reducing network traffic to the cloud by only uploading what is required [4] and performing computing tasks closer to the user. As fog nodes are closer to the end-user, users obtain a better-tailored application that suits their needs.

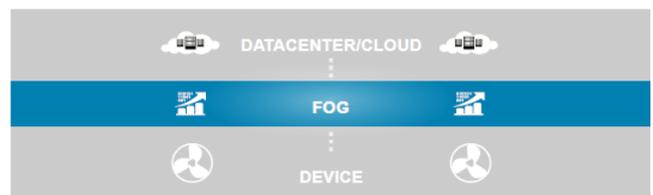

Figure 1. Fog lies between the Cloud and End-User (Taken from [4])

The structure of the paper is as follows. Section 2 describes some of the related concepts to fog computing, section 3 discusses a few of the computational algorithms that are currently in use, section 4 contains the discussion of the algorithms and the paper is concluded in section 5.

## II. RELATED CONCEPTS

The concept of fog computing is not a new one as there have been different relatable technologies that bring resources closer to the end user to improve processing, storage, and speed of applications. These include cyber foraging, cloudlets, and multi-access edge computing.

### A. Cyber Foraging

One of the first concepts used for edge computing was cyber foraging [3]. Making mobile devices small and light in weight while being able to perform desktop-level applications was a major challenge that needed to be addressed [5]. By creating a dynamic host of computing power and data staging servers nearby to a user, the needs of a wireless mobile computer can use these resources to perform applications that are limited to the mobile device [5]. Systems that are in place, that help mobile computers in computational and storage tasks, are called surrogates; where these systems are not administrated by an authority and are considered untrusted. For example, a mobile device will detect a surrogate in their operating environment and negotiate with the surrogate for their resources through short-range wireless communication technology and when the mobile device needs to process a relatively large amount of data, the mobile device directs this computation required to the surrogate [5]. For applications such as facial recognition, the mobile device will capture the images, pass it to the surrogate where it will perform the relevant processing algorithms using a database and deliver the result to the mobile device [3]. Information on the surrogate used for the mobile device is removed when the mobile device does not require the surrogate; or is disconnected from the surrogate. Data can also be staged by predicting what the user's needs to reduce cache miss service times [5]. As this is an effective solution to reduce latency and bandwidth constraints for a given user, challenges such as management, trust level and load balancing of surrogate devices need to be addressed.

### B. Cloudlet

Improving the concept of cyber foraging, cloudlets are trusted devices that are connected to the Internet which provide a large array of resources available to nearby mobile devices [6]. These are decentralized and dispersed across a large area for mobile computers to use. Cloudlets can be viewed as cloud-based systems on the edge of the network as they reuse cloud-based techniques such as virtual machines and are within a single-hop proximity of mobile devices [3]. Cloudlets act and perform as a scaled-down data center in a box that allows a mobile user to connect via Wi-Fi to create and initiate virtual machine instances [7]. If there are no cloudlet services around that a mobile device requires, the device can resort to using conventional cloud-based systems or its own dedicated hardware and software [6]. Looking again at facial recognition, processing will be performed on virtual machines instead of real machines which allow cloudlets to dynamically change depending on the intensity of processing required [3]. As cloudlets can represent full cloud infrastructures in certain scenarios, they can exist in a standalone environment due to virtual machines deployed without the resources of a conventional cloud [3].

### C. Multi-Access Edge Computing (MEC)

Also referred to as mobile edge computing, MEC focuses on mobile networks and virtual machines [3]. Working with real-time Radio Access Network (RAN) information, developers can create applications to suit a user's needs by obtaining data such as the user's location [8]. This information can also be useful to businesses wanted to sell a product or service by learning about a user's interests based on their location. By allowing resources to be located at the edge of the network, within the range of the RAN, the user's latency and bandwidth consumption decrease [8].

### D. Similarities and Differences with Fog Computing

All the proposed technologies have the same characteristics in common: to reduce latency, network constraints and provide the user with a product or service instantaneously. Cyber foraging is succeeded by cloudlets in terms of better security and management and can operate as a standalone device or connected to the cloud. Fog computing can operate in both standalone or connect to the cloud to help assist in applications and services. MEC has a fall back here in the case that it does not work with cloud-based systems and only in a standalone environment [3]. The main difference between the technologies is application usage. Cloudlets work by reducing processing and storage limits to a user whereas MEC works with applications that are best suited for edge environment operation [3]. Fog computing can take both characteristics by allowing compute and storage resources to lie both at the edge in a fog node, in the cloud or spanning across the two depending on the application requirements [3].

## III. COMPUTING ALGORITHMS

A general outline of how algorithms designed for fog systems have been outlined in [3] discussing points of heterogeneity, QoS management, scalability, mobility, federation, and interoperability. Heterogeneity is how fog components can be diverse in its application such as perform several different functions regardless of the environment placed in with respect to computation and storage needs. QoS management is how fog systems provide improved user experience by reducing latency and processing time. Scalability is how fog computing can be implemented on a great horizontal and vertical spectrum with regards to the number of users to the number of applications. This factor should be dynamic as there may not always be a need for a large number of fog components in low demand situations. Mobility is how fog components can be moved and relocated if necessary while in operating conditions to perform a function. This can be seen with vehicle-to-vehicle communication [9] where vehicles [10] can act as fog nodes used to collect data onboard a vehicle [11] while moving at different speeds and through different operating environments. Federation ties together with heterogeneity in the case that fog systems need to be diverse and be able to be managed by different operators with different requirements regardless of the fog components in place. Lastly, fog components need to be interoperable with other devices from other infrastructures such as cloud and IoT systems as fog helps improve these.

Fog infrastructure can be viewed from both algorithmic and architectural views. Algorithms for fog can be sub-divided into computing, content storage, and distribution, energy consumption and application-specific algorithms where architecture can be subdivided into application-agnostic architectures and application-specific architectures. As the field of fog computing is a very large one, this paper covers what has been done in the algorithmic environment with respect to computing in fog systems.

By analyzing and pairing nodes in a fog network, system efficiency can be improved by node pairing in the same fog domain [12] by using a utility-based scheme. This is done by refining Irving's matching algorithm and creating a model as a one-sided stable matching game with a quota [12]. A utility-based list is created that considers factors that include transmission power, transmission distance, and cost. A pairing of nodes is performed through the utilities by the creation of preference lists in each node to other nodes in the fog domain. By performing this matching algorithm for fog nodes in the same network, an improvement in performance in the form of full potential was seen [12]. By making use of a utility-based node matching algorithm, efficiency is improved over a greedy algorithm where a node would pair with a neighboring node without evaluation of the situation. By making use of short-range communication methods, local resources can be leveraged to reduce communication and computational loads.

By allowing for radio access clustering in fog computing applications, improved user experience is obtained by reducing power consumption or latency for each user [13]. By creating a small cluster of cells, the computation can be shared in a cluster to improve the efficiency of fog components and reduce network traffic. This is done by distributing the computational load across adjacent radio access points with computational capacities. Two approaches to creating clusters included forming a cluster to perform a task with the lowest latency and the other to reduce the amount of power usage [13]. Clusters contain inside communication and computational resource allocation and are scalable. Trade-offs can be seen by tuning the latency performance of a cluster by scaling the size of the cluster accordingly. This implementation shows an improvement over the algorithm used in [12] as this accommodates heterogeneity as well as QoS by reducing latency.

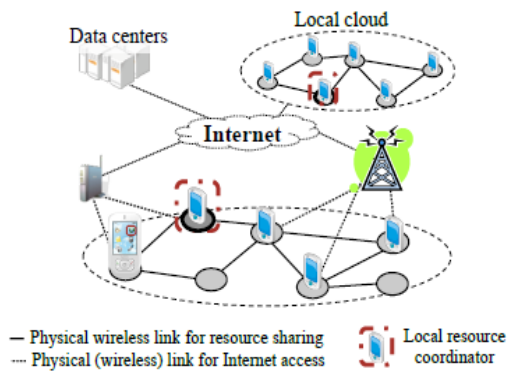

Figure 2. System model used for local resource coordinator (Taken from [14])

By mapping particular parameters to time resources, a mathematical framework for the sharing of resources is implemented [14]. The main issue is how to distribute resources between nodes in a fog system. A simple approach is to perform numerical calculations or downloading of data, however, this does not factor in the service that is being delivered. This can result in an increase in factors such as latency for latency intensive applications such as augmented reality. Therefore, by performing task-orientated sharing, the processing is minimized. This is done by creating a unified framework to map resources that include power, bandwidth, and latency to distribute resources. From these parameters, optimization problems are created and solved through convex optimization approaches.

From Fig. 2, the system model is shown where a local resource coordinator handles the resource sharing of the fog domain. Nodes in the fog domain communicate with each other through various short-range communication methods that include Wi-Fi and/or Bluetooth.

The local resource coordinator is selected by the nodes in the domain through factors that include connectivity, CPU performance, and energy usage. The coordinator receives messages from the connected nodes in the domain about the amount of resources it has and the tasks it needs to perform. The coordinator then assigns tasks to each node to maximize the resource usage of each node. This proves to reduce latencies as well as improve energy efficiency. Increasing the number of nodes in a fog domain enables the system to perform better due to more resources being available, however, the communication range between nodes should be increased which will result in an increase in energy usage [15, 16]. Therefore, as a constraint, resource sharing can improve both latency and energy usage to a certain degree.

Using a mathematical model, [17] makes use of defined policies that are constructed around latency used to determine where tasks are performed in a fog infrastructure. The problem definition is to create a low latency network that consists of a group of users in the same domain, acting as one source node and a fog server as another node. Three different policies are created that are used to determine which fog node is to perform a task. The first policy assigns a task to a randomly selected node from a uniform distribution. The second policy selects the fog node with the lowest latency given the current state of the fog system. The third policy makes use of the fog node with the largest amount of resources available to perform a task. The system determines which fog nodes are to be used in the candidate list for redistributing the workload based on the workload generated by each source node. Based on the policies, a fog node will process the workload. By adjusting the number of nodes available and threshold values, it was proven that the lowest latency policy was the most effective.

By looking at a service orientated resource management model, [18] makes use of a user's behavior towards the usage of a resource in order to allocate resources accordingly. This makes use of a customer's historical use of the resources over time. Usage of resources can become more efficient when there is a predictive model in place to determine what resource needs to be used. With new users, default probabilities are used which results in the system. This can prove to be an issue if a new user takes advantage of resources over a user that has a good history for use of resources. An issue with this approach is that scheduling of tasks does not consider the heterogeneity of fog components which can degrade the performance of the fog system. Considering the type of service, overall service relinquishes probability and service-oriented relinquish probability, this implementation helps improve resource management that prevents wasting resources.

Making use of edge intelligence in fog computing, [19] makes use of two implementations that will help with application offloading and storage expansion for mobile devices [20, 21]. With application offloading, resources that are viewed include network latency, bandwidth, size of the overhead data and dynamic allocation of the partitions used for the loading. In storage expansion, making use of the user's devices that are near is viewed to improve latency, security, and privacy. Linear regression models usually result in static decisions since they do not account for the dynamic environment of resources being used. By monitoring resources constantly, intelligence in fog computing can encourage effective offloading decisions in a fog network. A test bench of 5 different environments is created with three representing fog-based systems and two representing cloud-based systems. The fog systems contain short-range communication methods such as WLAN or Wi-Fi which help reduce latency, opposed to the cloud systems that make use of 3G or 4G connectivity while cloud systems contain better CPU performance over the fog systems.

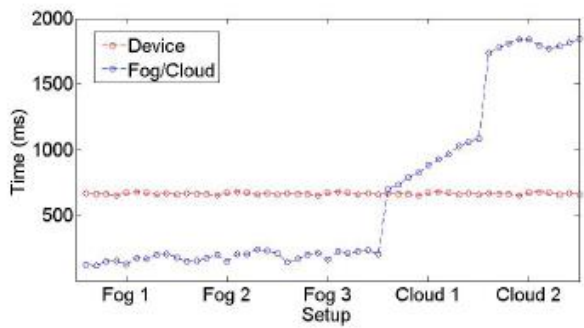

Figure 4. Fog vs Cloud execution time (Taken from [14])

Making use of the face recognition method of Picaso, it can be viewed in Fig. 3 that the fog systems reduce the time drastically to process data. From this result, a model can be created to predict the performance of various tasks to offload them efficiently based on parameters such as bandwidth, latency, CPU usage, and memory. For the offloading of storage, the goal is to achieve a reduction in cost and latency for the user. This is done through a geographically distributed location policy that creates a cluster of files based on the data type, operation, access and access in the future [19].

Using Cloud Atomization Technology, fog nodes on different levels can be used to create virtual machine nodes [22]. This is performed using graph partitioning to improve load balancing techniques in a fog system. By doing this, a dynamic load balancing method is effectively implemented to handle resources and reduce the consumption of node migration that is introduced by system changes [22]. This implementation is focused on a balancing algorithm that is proportional to the number of resources available. Improvement of this algorithm can be done by analyzing the change in the system environment

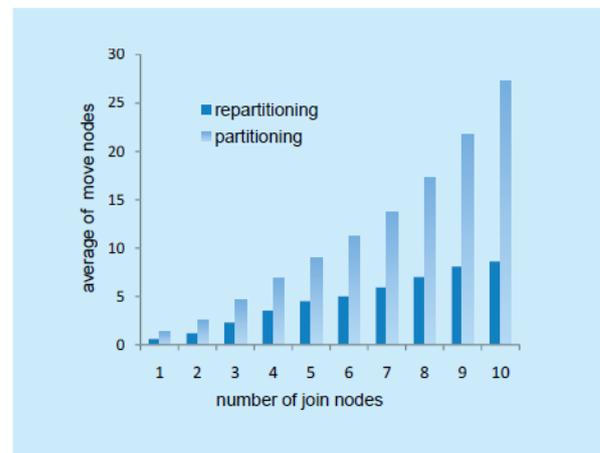

Figure 3. Cloud atomization using graph partitioning scheme (Taken from [22])

and improving the performance of the elastic load balancing method.

Looking at Fig. 4, the number of nodes that are moved is considerably less when repartitioning the fog system for load distribution. Making use of a static balancing algorithm will need to assess the load of each node which can result in partitioning occurring more often than required to balance the workload. By repartitioning, there

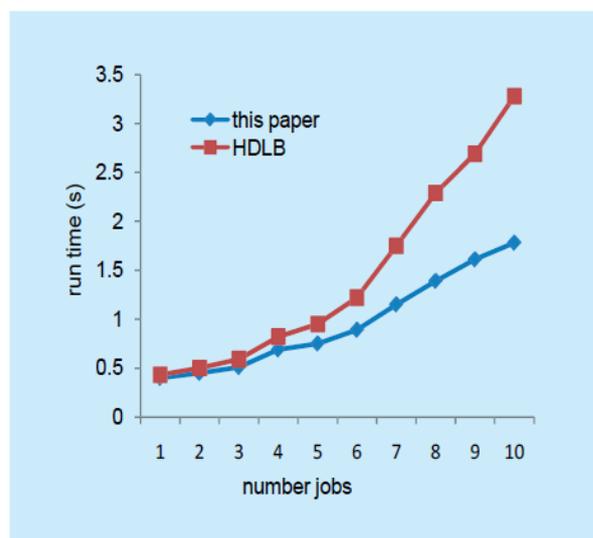

is a reduction in the number of resources used and waiting time.

Figure 5. Cloud atomization repartitioning scheme run time (Taken from [22])

Fig. 5 shows a reduced run time of the repartitioning scheme compared to the classical hybrid strategy. Mentioned earlier, the workload will need to be computed each time before a partition can take place which will increase the runtime of the system.

Reducing redundant elements of computing and storage elements in a fog system can reduce the bandwidth consumption and latency for users [23]. Making use of coding concepts that include minimum bandwidth codes and minimum latency codes. Minimum bandwidth codes create a trade-off between computational and communication loads by making use of computing resources that are not being used to their potential to reduce communication loads. Minimum latency codes are the opposite of minimum bandwidth codes in the sense that applications can be completed much faster by incorporating more computational resources. This can be further improved by making use of a unified coding framework that decreases the communicational and computational load of a system.

## IV. DISCUSSION

All the discussed computing algorithms in section III provide a better understanding as to what can be achieved from an algorithmic point of view for improving fog systems. The system proposed in [9] created an efficient way to utilize the maximum potential of resources in a short-range communication spectrum. This gives the heterogeneity of quality to the system. This system is improved by adding for QoS through dynamic sizing and management of resources of the computation clusters [10]. This is also the same case with [11], [13], [16] where the fog system is heterogeneous and QoS is achieved. QoS is achieved only with [12], [14] and [15]. What is the challenge in these implementations is the scalability and mobility factors. Considering the characteristics of what fog systems should contain, these algorithms lack federation and mobility. This is where fog algorithms can be further developed, by incorporating all the factors of heterogeneity, QoS, scalability, mobility, and federation to create a tradeoff system to ultimately improve cloud and fog systems.

## V. CONLUSION

This paper described what fog computing is, the challenges that exist in this infrastructure and what can be obtained through algorithms that are designed for fog systems. These solutions can also be applied to cloud-based systems where applicable to improve computing, network bandwidth, and latency as well as storage constraints.